%% file: main.tex
\documentclass[journal=ancac3,layout=twocolumn,manuscript=article]{achemso}
\makeatletter
\AtBeginDocument{%
  \@ifpackageloaded{natmove}{%
    \let\@citex\nmv@citex@nat
    \let\cite\nmv@cite
  }{}%
}
\makeatother

\usepackage[fontsize=9pt]{scrextend}

\usepackage{geometry}
\usepackage{charter}

\usepackage{type1cm}
\usepackage{lettrine}

\usepackage{xcolor}
\definecolor{ncomorange}{RGB}{255,102,0}
\usepackage{graphicx}
\usepackage{placeins}
\usepackage{siunitx}
\usepackage{todonotes}
\usepackage{hyperref}
\mciteErrorOnUnknownfalse

\let\oldmaketitle\maketitle
\let\maketitle\relax

\author{Chenyang Zhao}
\affiliation{Cavendish Laboratory, University of Cambridge, JJ Thomson Avenue, Cambridge CB3 0US, United Kingdom}
\email{cz390@cam.ac.uk}
\author{Sam M. Lambrick}
\affiliation{Cavendish Laboratory, University of Cambridge, JJ Thomson Avenue, Cambridge CB3 0US, United Kingdom}
\alsoaffiliation{ISIS Facility, Rutherford Appleton Laboratory, Chilton, Didcot, Oxfordshire OX11 0QX, United Kingdom}
\email{sml59@cantab.ac.uk}
\author{Ke Wang}
\author{Shaoliang Guan}
\author{Aleksandar Radi\'{c}}
\author{David J. Ward}
\author{Andrew P. Jardine}
\email{apj24@cam.ac.uk}
\author{Boyao Liu}
\email{bl476@cam.ac.uk}
\affiliation{Cavendish Laboratory, University of Cambridge, JJ Thomson Avenue, Cambridge CB3 0US, United Kingdom}

\date{\today}

\title{Characterising Atomic-Scale Surface Disorder on 2D Materials Using Neutral Atoms}

\begin{document}
\twocolumn[
\begin{@twocolumnfalse}
\vspace*{-1cm}
\oldmaketitle
\vspace*{-0.4cm}
{{\rule{\textwidth}{1pt}}}

\vspace*{0.2cm}
\textbf{Two-dimensional (2D) transition metal dichalcogenides (TMDs), such as MoS\textsubscript{2}, have the potential to be widely used in electronic devices and sensors due to their high carrier mobility and tunable band structure. In 2D TMD devices, surface and interface cleanness can critically impact the performance and reproducibility. Even sample surfaces prepared under ultra-high vacuum (UHV) can be contaminated, causing disorder. On such samples, trace levels of submonolayer contamination remain largely overlooked, and conventional surface characterisation techniques have limited capability in detecting such adsorbates. Here, we apply scanning helium microscopy (SHeM), a non-destructive and ultra-sensitive technique, to investigate the surface cleanness of 2D MoS\textsubscript{2}. Our measurements reveal that even minute amounts of adventitious carbon induce atomic-scale disorder across MoS\textsubscript{2} surfaces, leading to the disappearance of helium diffraction. By tracking helium reflectivity over time, we quantify the decay of surface order across different microscopic regions and find that flat areas are more susceptible to contamination than regions near edges. These findings highlight the fragility of surface order in 2D materials, even under UHV, and establish SHeM as a powerful tool for non-damaging microscopic 2D material cleanness characterisation.  The approach offers a new route to wafer-scale characterisation of 2D material quality.}

{{\rule{\textwidth}{1pt}}}
\vspace*{0.4cm}
\end{@twocolumnfalse}
]

\section{Introduction}
Two-dimensional transition metal dichalcogenides (2D TMDs), such as molybdenum disulfide (MoS\textsubscript{2}), have emerged as semiconductor materials for next-generation electronic and optoelectronic devices due to their atomically thin structure, tunable bandgap, and high carrier mobility \cite{Manzeli2017}. The properties of TMDs are highly sensitive to changes to surface conditions. Specifically, surface contamination consisting of chemisorbed and physisorbed species has an effect on the electronic properties of such 2D materials\cite{Kretinin2014,Hou2023,Park2023}, and is known to affect the performance of nanoscale devices \cite{Jariwala2013,Tongay2013}. The influence of surface contaminants is significantly more acute than for traditional silicon semiconductors, where the electronic properties are determined by the bulk material. For 2D materials, contamination layers can be as thick as the semiconductor itself and it is often the reduced dimensionality that provides the advantageous electronic properties of these materials. Thus surface cleanness characterisation has become a critical concern in both fundamental studies and practical device applications \cite{Wu2024,Wang2022,Illarionov2020,Dong2024,Sovizi2023,Wu2022}.
However, weakly bound adsorbates, often referred to as being physisorbed, have proven difficult to identify and measure, especially on wafer scale systems. Even if devices are to be encapsulated, the presence of contamination between layers during fabrication remains an important issue.  Indeed, on the wafer scale contamination is a greater concern than on the nanoscale, as the ``self-healing'' processes \cite{Dong2024,Haigh2012,Rooney2017,Kretinin2014} observed with devices constructed by mechanical exfoliation may not be applicable over large areas.  

To address the challenges of detecting weakly bound surface contaminants, we focus on adventitious carbon, a ubiquitous contaminant even under UHV conditions\cite{Grey2024,Greczynski2024,Barr1995,Greczynski2020}, on MoS\textsubscript{2}, a prototypical 2D semiconductor actively being utilised in device applications\cite{Manzeli2017,Guros2019,Chen2021}.  We introduce a novel approach for characterising weakly bound surface adsorbates over large sample areas by applying scanning helium microscopy (SHeM) \cite{barr_design_2014,witham_simple_2011,Palau2023,Zhao2025}, a technique which uses low energy neutral helium atoms to provide exceptional sensitivity to surface contamination.

Adventitious carbon, as a predominantly aliphatic hydrocarbon-based contaminant with minor oxygen functionalities \cite{Biesinger2022,Grey2024}, significantly affects the chemical, structural, and electronic properties of 2D materials. For example, it modifies surface chemistry and energetics, causing increased hydrophobicity \cite{Kozbial2015}, frictional anisotropy \cite{Pálinkás2022}, and altered work function or contact properties \cite{Chen2021,Kim2024}. In the electronic regime, adventitious carbon introduces electronic disorder and charge inhomogeneity \cite{Kretinin2014,Hou2023}, shifts doping behaviour \cite{Hou2023,Park2023}, and contributes to mobility degradation in devices \cite{Kretinin2014,Chen2021,Guros2019}. These impacts extend further to interface integrity and device reproducibility. Even trace adventitious carbon residues can disturb interfacial crystallinity \cite{Bayer2018} and contribute to performance variability across nominally identical devices \cite{Guros2019}. In some cases, adventitious carbon can act as a reactive site for oxidation and etching \cite{Park2020}. Moreover, its spectral complexity and variability 
interfere with spectroscopic interpretation \cite{Barr1995,Marinov2023}. While annealing, sputtering or tip-based cleaning can effectively remove adventitious carbon \cite{Guros2019,Bayer2018,Kim2024,Buitrago2024,Chen2021}, its re-adsorption under UHV conditions is not well characterised and has attracted limited attention in the 2D materials community.

Previous characterisation of contaminants on 2D/van der Waals (vdW) materials has generally followed two directions. First, X-ray photoelectron spectroscopy (XPS) has been used to identify the chemical composition of the contamination \cite{Greczynski2020,Addou2015}. However, traditional XPS does not possess spatial resolution, and selected-area XPS \cite{bozzini_spatially_2019} has reduced sensitivity, which is crucial for thin layers of contaminant. Second, the morphology of the contamination has been explored with scanning tunnelling microscopy (STM) \cite{Pálinkás2022} and transmission electron microscopy (TEM) \cite{Park2020}. These two approaches represent extremes in the length scale, with XPS generally sensitive to macroscopic areas, while TEM and STM are restricted to very small sample areas. Neither current approach represents a practical method of testing surface cleanliness over both larger areas (wafer scale) and with spatial resolution. In addition, both TEM and XPS use high energy probes, which have sufficient energy to cause desorption as well as changes/damage to the sample \cite{jain_adatom-mediated_2024,krishna_review_2022}. STM is generally considered non-damaging, however, even then tip-induced desorption is a known phenomenon \cite{moller_automated_2017}. Addressing the characterisation gap is essential for improving the consistency and fidelity of device fabrication \cite{Wu2024,Wang2022,Illarionov2020,Dong2024,Sovizi2023,Wu2022,Haigh2012,He2025,Zhao2025nce}.

In the current work we present an alternative method. We utilise SHeM \cite{barr_design_2014,witham_simple_2011,Palau2023,Zhao2025} to detect the presence of contamination over sample areas of multiple mm and with spatial resolution down to the $\SI{}{\micro\metre}$ scale. Further, we use helium atom micro-diffraction (HAMD) \cite{vonJeinsen2023,hatchwell_measuring_2024} and neutral helium imaging with SHeM to reveal that MoS\textsubscript{2} surfaces under UHV lose atomic-scale order due to trace-level adsorption of adventitious carbon. The energy of the incident helium atoms used (thermal energy, $\sim\SI{64}{meV}$) means the incident beam cannot induce changes on the sample, such as desorption, that can occur with energetic beam techniques such as XPS and TEM. Thus, SHeM is especially suited for weakly bonded adsorbates tricky to detect with other techniques. The neutral atom-scattering process is also chemically agnostic, so our approach can be applied to many different potential sources of contamination.

In addition, we dynamically record the time-decay of the surface order with microscopic resolution, to show the process of re-adsorption of contamination, even 
under UHV conditions. We also demonstrate that the surface contamination of adventitious carbon on MoS\textsubscript{2} can be removed by heating to as little as $\ang{80}$C,
significantly lower than previously reported \cite{Addou2015}. The process of cleaning and re-contamination is found to be repeatable, and the chemical nature of the contaminants is confirmed with XPS. We study two regions of our sample that exhibit different levels of crystallinity, and find that regions with larger areas of crystallinity re-contaminate faster than those that exhibit only local regions of crystallinity. 

\section{Helium atoms as probes of surface contamination}

\begin{figure}[b]
    \centering
    \includegraphics[width=\linewidth]{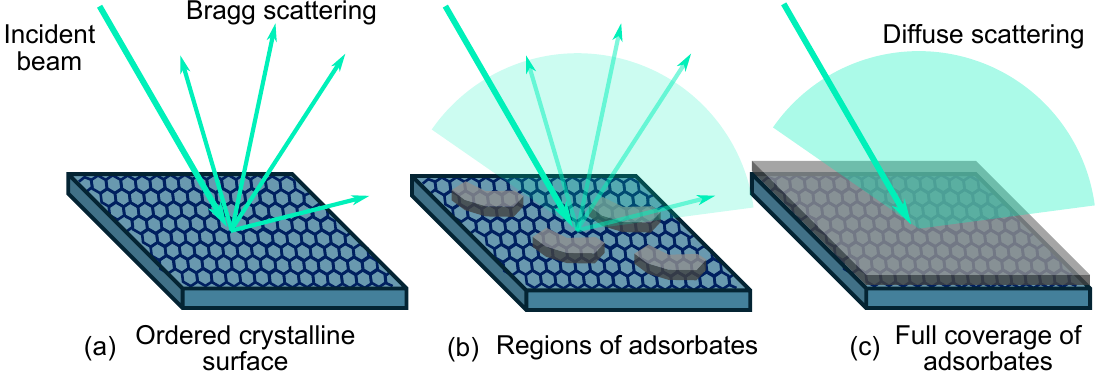}
    \caption{\textbf{Schematics of helium atom diffraction from crystalline surfaces under different levels of contamination.} (a) For an ordered crystalline surface, such as the surface of an MoS\textsubscript{2} crystal, helium atoms scatter undergo Bragg scattering (ordered scattering). (b) In the presence of adsorbates, the atoms are scattered randomly, reducing the intensity of ordered scattering and increasing the intensity of disordered diffuse scattering. (c) A full coverage of disordered adsorbates results in completely diffuse scattering.}
    \label{fig:surface_contamination}
\end{figure}

SHeM is a surface characterisation technique, that uses the scattering of a focused or collimated  ($\sim0.1-\SI{10}{\micro\metre}$ dia.) low-energy beam of neutral helium atoms ($\sim$ 64 meV incident energy) to probe the top layer of a sample pixel by pixel. The low-energy of the probe atoms provides ultra-high surface sensitivity, with helium atoms scattered by the valence electrons in the top atomic layer of the surface \cite{holst_material_2021}. The low energy of the atoms also results in a de Broglie wavelength commensurate with inter-atomic spacings, allowing SHeM to probe atomic-scale surface order via diffraction on a microscopic length scale \cite{vonJeinsen2023,radic_defect_2024}. In particular, helium scattering is known to be sensitive to surface adsorbates, even in low coverages, due to a large cross section to atomic-scale defects \cite{farias_atomic_1998}. Recent work with SHeM has shown high sensitivity to vacancy-type defects \cite{radic_defect_2024}, with the ratio of ordered to disordered scattering providing a quantification of the coverage.

Surface contamination is measured in SHeM by observing the level of order in the scattering of the probe helium atoms from the surface. For an ordered crystalline surface, helium atoms will undergo scattering following Bragg's law, producing diffraction peaks in the scattering distribution, as shown in figure \ref{fig:surface_contamination}(a). In the presence of a submonolayer of adsorbates, the symmetry of the surface is partially broken, causing some of the intensity to be scattered diffusely, as seen in figure \ref{fig:surface_contamination}(b) (the distribution of the diffuse scattering is often assumed to be cosine-like \cite{Lambrick2022}, a model associated with completely random scattering). If an entire layer of disordered adsorbates forms on the surface, as shown in figure \ref{fig:surface_contamination}(c), then all of the scattering will be diffuse, and there will be no sharp peaks in the measured data.

In SHeM micrographs, the presence or lack of ordered Bragg scattering can be inferred via the presence of bright `shiny'  regions and rapid changes in the intensity in the absence of significant topography, these image features have been referred to as diffraction contrast in the literature \cite{vonJeinsen2023,Bergin2020,hatchwell_measuring_2024}. The diffraction peaks can be observed by performing a HAMD measurement, where the intensity is measured as a function of the scattering angle, which provides confirmation of the origin of the contrast observed in the associated micrographs.

The desorption, and re-adsorption, of surface contaminants can be monitored, for a microscopic spot on the sample, by monitoring the intensity of the specular peak while the temperature of the sample is modified. The term helium reflectivity is often used for the relative intensities of the specular signal. An increase in the  observed intensity with increasing temperature indicates surface containments leaving the surface. Recontamination can be inferred by an observed decrease in reflectivity when the temperature is lowered. Further experimental details can be found in the Methods section, with additional description in a recent publication on the Cambridge `B-SHeM' \cite{Zhao2025}, which was used for data acquisition in the current work.

\section{Results}

A crystal of natural MoS\textsubscript{2} was used as our primary test candidate. The crystal was cleaved with scotch tape prior to loading into the SHeM sample chamber. To confirm our results with single layer MoS\textsubscript{2} we prepared a monolayer flake of MoS\textsubscript{2} on top of a hexagonal boron nitride (hBN) buffer on a SiO\textsubscript{2} substrate. 

\begin{figure*}
    \centering
    \includegraphics[width=0.65\linewidth]{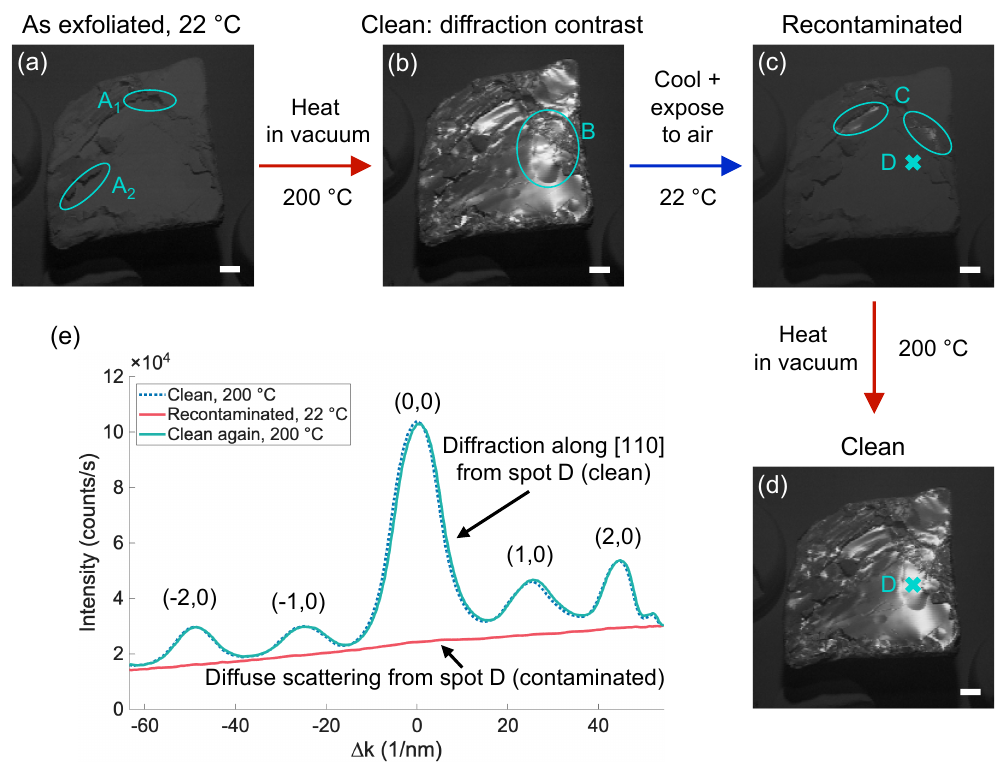}
    \caption{\textbf{SHeM measurements of MoS\textsubscript{2} under different levels of contamination.} (a-d) SHeM image of the sample as exfoliated (a), after heating to $\SI{200}{\degreeCelsius}$ for an hour (b), after cooling back down to room temperature and exposure to air for $\sim\SI{30}{mins}$ (c), and after heating again to $\SI{200}{\degreeCelsius}$ for an hour (d). The scale bar is $\SI{500}{\micro\metre}$. (e) HAMD measurements from spot D on the clean sample surface at $\SI{200}{\degreeCelsius}$ and contaminated surface at room temperature. (a) exhibits only diffuse contrast, with slowly varying intensity, (b) exhibits diffraction contrast, with many bright spots and rapidly changing intensity even for relatively flat regions, (c) exhibits mostly diffuse contrast with some regions of diffraction evident, (d) exhibits the same contrast as (b).}
    \label{fig:clean_and_recontaminate}
\end{figure*}

\begin{figure}[t]
    \includegraphics[width=0.8\linewidth]{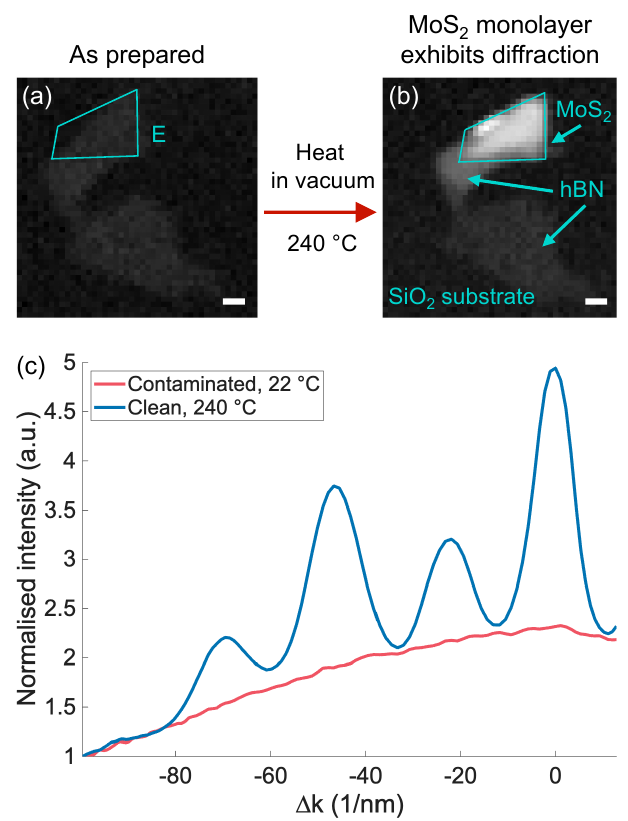}
    \caption{\textbf{SHeM measurements of monolayer MoS\textsubscript{2}.} (a-b) A monolayer of MoS\textsubscript{2}-on-hBN as prepared in air (a), and after heating in vacuum (b). The monolayer was identified by optical microscopy and is highlighted as region E, the remaining bright region in (b) is the hBN buffer. The scale bar is $\SI{10}{\micro\metre}$. (c) Diffraction scans from the monolayer MoS\textsubscript{2} at room temperature and $\SI{240}{\degreeCelsius}$. The same change in contrast is seen as with the bulk sample in figure \ref{fig:clean_and_recontaminate}, with the distinction between the MoS\textsubscript{2} crystal and the substrate barely visible prior to cleaning, but with the crystal exhibiting high reflectivity after cleaning. All measurements of monolayer samples were taken with the Cambridge A-SHeM \cite{barr_design_2014,vonJeinsen2023}.} 
    \label{fig:monolayer_mos2}
\end{figure}

Figure \ref{fig:clean_and_recontaminate} shows measurements during the process of cleaning the MoS\textsubscript{2} to remove adsorbates, then their reabsorption upon cooling. Upon initial measurements, (a), the crystal exhibits diffuse topographic contrast, with only slowly varying intensity and the strongest features being due to masking of the detection direction by surface topography (highlighted as feature A). Diffuse scattering implies disorder on the surface at the atomic scale, as in figure \ref{fig:surface_contamination}(c). Heating the sample to $\SI{200}{\degreeCelsius}$ for $\sim\SI{1}{hour}$ leads to a change in the observed contrast in figure \ref{fig:clean_and_recontaminate}(b), where we see bright highly reflective regions, along with rapid changes in contrast (highlighted as feature B), while the masks that previously dominated contrast are now only minor features. These contrast features imply diffraction contrast \cite{Bergin2020,vonJeinsen2023}, and hence ordered Bragg scattering. We attribute the observed disorder in figure \ref{fig:clean_and_recontaminate}(a) to a layer of adventitious carbon that formed between exfoliation and imaging, which is further confirmed by XPS measurements on the same sample (see the supplementary information for the XPS results).

To demonstrate the repeatability of the process, the sample was re-exposed to ambient air for $\sim$ 30 minutes, the sample chamber was then evacuated and the sample stayed in the vacuum chamber for $\sim$ 24 hours at room temperature prior to re-measuring. When re-imaged (figure \ref{fig:surface_contamination}(c)), most of the surface again exhibited diffuse contrast and hence were covered with contamination, although small bright regions remain clean (highlighted as feature C). When the sample was heated again to $\SI{200}{\degreeCelsius}$, the MoS\textsubscript{2} crystal surface returned to a clean state, \ref{fig:clean_and_recontaminate}(d), with similar contrast features to (b). In addition, HAMD patterns in figure \ref{fig:clean_and_recontaminate}(e) (data acquired from spot D), confirms our interpretation of the image contrast as arising from diffraction and diffuse scattering for clean \& ordered surfaces and contaminated \& disordered surfaces, respectively. The reversible adsorption and desorption behaviour of surface contaminants is confirmed by the changes observed in figure \ref{fig:clean_and_recontaminate}(a)$\rightarrow$(d).


Further to our investigations with a bulk MoS\textsubscript{2} crystal, we prepared a monolayer MoS\textsubscript{2} sample, on which we observed similar behaviour. A sample of monolayer MoS\textsubscript{2} was prepared by mechanical exfoliation and placed on top of a $\sim\SI{25}{\nano\metre}$ thick hBN buffer. The hBN protects the morphological
and electronic properties of the monolayer MoS\textsubscript{2}, as previously shown using low-energy electron microscopy/diffraction and photoluminescence \cite{Man2016ProtectingBuffer}.
Figure \ref{fig:monolayer_mos2} shows a prepared flake of monolayer MoS\textsubscript{2} (highlighted as feature E and identified as the monolayer flake with optical microscopy) imaged first at room temperature (a), then after heating to $\SI{240}{\degreeCelsius}$ for a few hours (b), and diffraction measurements from the monolayer MoS\textsubscript{2} before and after heating (c). We observed the same change in contrast as with the bulk sample, initially almost no distinction can be made between the MoS\textsubscript{2} and the diffusely scattering substrate, but after heating the MoS\textsubscript{2} starts to exhibit diffraction, demonstrating an atomically clean surface. The consistency between both monolayer and bulk MoS\textsubscript{2} results enables conclusions drawn from studies of the bulk surface to be applied to atomically thin regimes.

To explore the recontamination process under UHV (sample chamber $\sim2\times10^{-8}\ \SI{}{mbar}$), the MoS\textsubscript{2} crystal was cleaned via heating to $\SI{200}{\degreeCelsius}$, then allowed to cool to $\SI{22}{\degreeCelsius}$. After cooling the specular reflectivity was monitored as a function of time, allowing us to quantify changes in atomic scale surface order. Two regions of the sample were studied, first a relatively flat region, corresponding to the bottom right of the micrographs in figure \ref{fig:clean_and_recontaminate}, representing a surface with larger-scale crystallinity. Second, a region of the sample that had delaminated during exfoliation was examined, representing a surface with only smaller-scale crystallinity, corresponding to the top right of the micrograph in figure \ref{fig:clean_and_recontaminate}.

\begin{figure*}[]
    \centering
    \includegraphics[width=\linewidth]{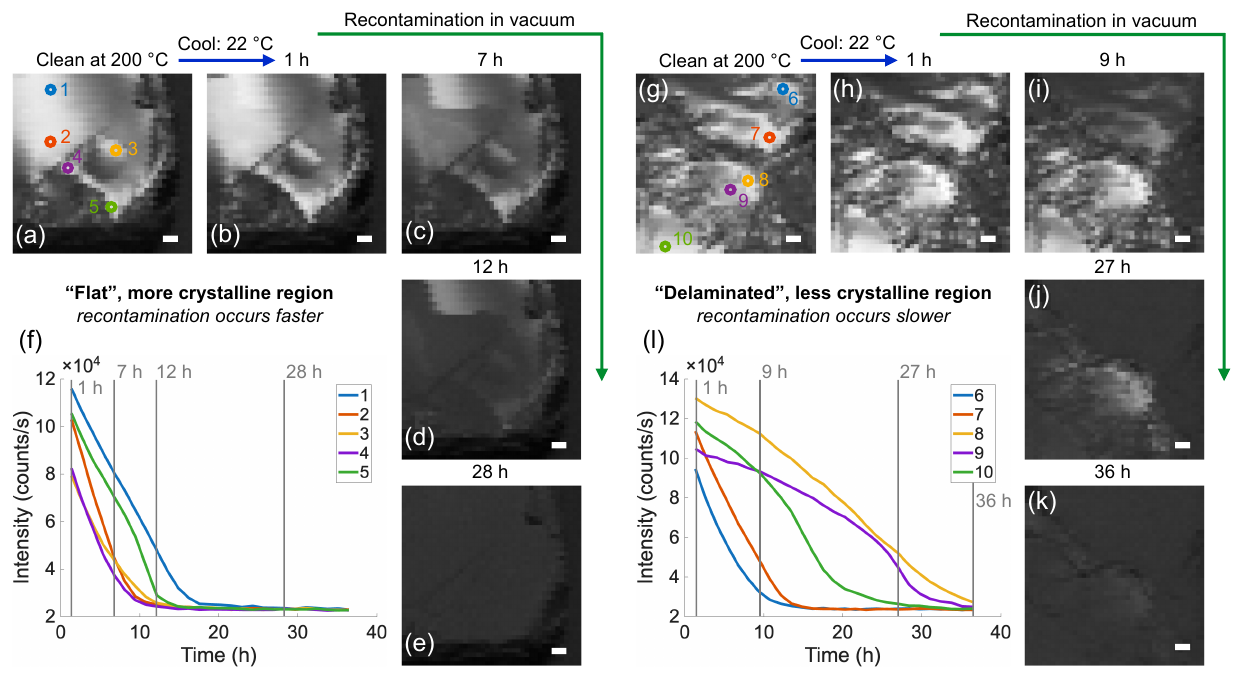}
    \caption{\textbf{Spatial resolution of the contamination process on MoS\textsubscript{2}.} Two regions of the MoS\textsubscript{2} crystal from figure \ref{fig:clean_and_recontaminate} are shown, one corresponding to a generally flat region (left, panels (a) to (f)) and the other a delaminated region (right, panels (g) to (l)).
    After cleaning the sample surface at a high temperature, (a) \& (g), diffractive contrast was observed. The sample was cooled to room temperature while remaining under vacuum and the specular signal monitored for 5 different points on each sample (f) \& (l). At intervals, micrographs of both regions were acquired demonstrating the changing contrast, (b)-(e) \& (h)-(k). Through both the specular intensity and contrast in micrographs it was observed that surface order remained for longer on the `delaminated' regions than the `flat' regions, and that complete recontamination was observed for all regions despite the sample remaining under UHV conditions. Scale bar, $\SI{100}{\micro\metre}$. As with the results in figures \ref{fig:clean_and_recontaminate} and \ref{fig:monolayer_mos2}, we also performed helium atom diffraction measurements to confirm our interpretation of the observed contrast, these are given in the supplementary information.}
    \label{fig:recontamination}
\end{figure*}

Figure \ref{fig:recontamination} presents helium micrographs and reflectivity measurements of the MoS\textsubscript{2} surface, demonstrating recontamination via adsorption of contaminants. Data in the left group of panels is from a relatively flat region of the sample exhibiting crystallinity over a larger sample area. Data in the right group of panels is for a delaminated region of the sample that exhibits poorer long range order, but that still possesses crystallinity over smaller length scales. Initial micrographs of both regions show clear diffraction contrast; subfigures (a) \& (g), indicating a highly ordered MoS\textsubscript{2} surface and function as references for high surface cleanness. After initial cooling, five points in each region of the sample were monitored over time.  In both regions the reflectivity then dropped over time, (f) \& (l), indicating a reduction in surface order.  The reduction was generally faster in the flat region of the sample. As the observed specular intensity dropped, the contrast in micrographs, taken every few hours, was seen to change; (b)-(e) for the flat region, (h)-(k) for the delaminated region. Initially, both regions exhibit diffraction contrast, which transitions into diffuse topographic contrast. After 28 hours and 36 hours respectively for the two regions, the observed contrast became the same as after exposure to air (figure \ref{fig:clean_and_recontaminate}(a)). As with the reflectivity, we observe that some of the delaminated region exhibits diffraction contrast for significantly longer than the flat region. The rate of contamination for the delaminated region was up to $\times 3$ slower than the flat region (spots 3 \& 4 in figure \ref{fig:recontamination} plotted in yellow and purple). We propose from our observations that larger areas of sample, showing higher levels of large-scale crystallinity, are more susceptible to recontamination by adventitious carbon than those more disordered on the scale $\sim\SI{100}{\micro\meter}$ (although still ordered on the atomic-scale). One possible explanation is that the adventitious carbon gets between the MoS\textsubscript{2} layers in delaminated regions. However, we lack direct evidence to confirm this, and further investigation is needed.



As with the initial experiment demonstrating decontamination and recontamination (figure \ref{fig:clean_and_recontaminate} \& \ref{fig:monolayer_mos2}), a set of HAMD measurements of the MoS\textsubscript{2} crystal were acquired to confirm our interpretation of image contrast and the measured specular intensity.
We observe a
reduction in the fraction of exposed pristine surface, which corresponds to decreasing
diffraction intensity, representing decreasing atomic-scale order resulting from increased adsorption of contamination. Full details are provided in the supplementary information.

\begin{figure*}[htbp]
    \centering
    \includegraphics[width=0.7\linewidth]{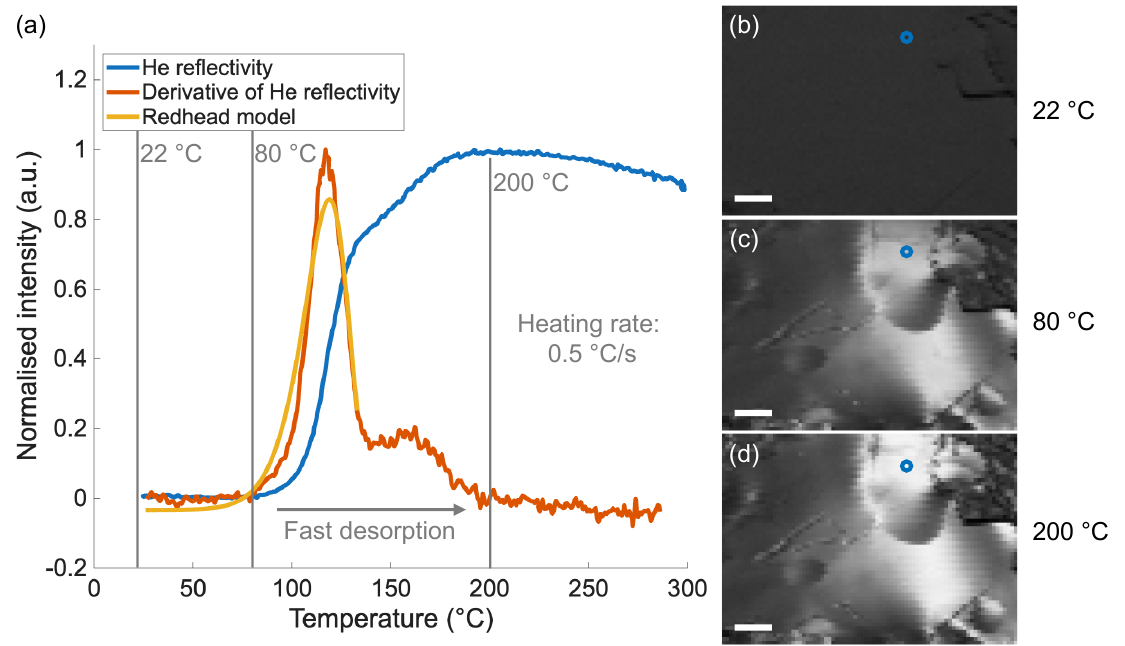}
    \caption{\textbf{Disappearance of contamination on MoS\textsubscript{2} by thermal heating.} (a) Temperature programmed helium reflectivity measurements of the MoS\textsubscript{2} crystal, with background subtracted. Helium reflectivity starts to increase at $\sim\SI{80}{\degreeCelsius}$ due to the desorption of adventitious carbon, reaches a maximum increase rate at $\sim\SI{120}{\degreeCelsius}$ and rolls off at $\sim\SI{200}{\degreeCelsius}$. (b)-(d) Helium micrographs of the bottom right corner of the MoS\textsubscript{2} sample in figure \ref{fig:clean_and_recontaminate} at room temperature, $\SI{80}{\degreeCelsius}$ and $\SI{200}{\degreeCelsius}$, imaged separately to the data in (a). We already observe diffraction contrast across the sample at $\SI{80}{\degreeCelsius}$, with no significant change when the temperature is increased to $\SI{200}{\degreeCelsius}$. The blue spot is the same as spot D in figure \ref{fig:clean_and_recontaminate}, where the helium reflectivity is measured. Scale bar, $\SI{300}{\micro\metre}$.}
    \label{fig:desorption}
\end{figure*}

To assess the desorption mechanism, temperature-dependent SHeM imaging and helium reflectivity measurements were performed on the same MoS\textsubscript{2} crystal, starting with the surface in a contaminated state. Temperature programmed helium reflectivity measurements of the MoS\textsubscript{2} after air exposure are shown in figure \ref{fig:desorption}(a). 
The reflectivity begins to rise around $\SI{80}{\degreeCelsius}$, corresponding to the thermal activation of adsorbate desorption. The increase in rate peaks near $\SI{120}{\degreeCelsius}$ and the reflectivity gradually levels off by $\sim\SI{200}{\degreeCelsius}$. The activation energy is estimated to be $\sim\SI{1}{eV}$ by assuming the derivative of the helium reflectivity behaves similarly to a conventional thermal desorption measurement, then fitting the derivative using the Redhead model \cite{Redhead1962} (See more details in the Supplementary Information). Between $\SI{200}{\degreeCelsius}$ and $\SI{300}{\degreeCelsius}$, the helium reflectivity slowly falls, a feature attributed to the Debye-Waller effect, which arises from increased thermal vibrations of surface atoms. In a separate contamination cycle, micrographs were taken of the bottom right region of the sample in figure \ref{fig:clean_and_recontaminate}; first at room temperature, figure \ref{fig:desorption}(b), then at $\SI{80}{\degreeCelsius}$, (c), which the reflectivity measurements suggest is the onset of desorption during the experiment, then finally at $\SI{200}{\degreeCelsius}$, (d), where the MoS\textsubscript{2} surface is known to exhibit diffraction in SHeM. At room temperature, diffuse topographic contrast is observed. Upon gentle heating to $\SI{80}{\degreeCelsius}$ for $\sim$ 10 minutes, diffraction contrast is observed. Upon further heating, the contrast is not observed to change, therefore we conclude that the surface was already removed of adventitious carbon at $\SI{80}{\degreeCelsius}$. These measurements confirm that the adventitious carbon can be removed by gentle heating in an ultra-high vacuum chamber, and MoS\textsubscript{2} surfaces can remain clean at the relatively modest temperature of $\sim\SI{80}{\degreeCelsius}$, significantly lower than the annealing used in previous work \cite{Addou2015,Kim2024,Buitrago2024,Bayer2018}.

\section{Conclusion}

We have introduced a novel method of identifying and characterising lightly bound adsorbates on the surface of vdW/2D materials, that can be applied to wafer scale areas.
Our technique uses the scattering of neutral helium atoms, which is highly sensitive to atomic-scale order, via a scanning helium microscope.  The method allows us to directly visualise the temporal evolution of microscopic surface cleanness with ultra-high sensitivity and no damage or beam-induced desorption. As an example system, we studied adventitious carbon on MoS\textsubscript{2}, a ubiquitous surface contaminant on TMD materials that is known to impact electronic properties and device performance.

Our results demonstrated that adventitious carbon was present on cleaved MoS\textsubscript{2}, and confirmed that it can be removed with heating \emph{in vacuo}. In addition, we showed that the carbon reappears after a brief exposure to air ($\sim\SI{30}{min}$), and also reappears under UHV conditions over a time scale of 10 to 30 hours. Further, our measurements revealed that regions of MoS\textsubscript{2} exhibiting larger-scale crystallinity re-contaminate faster under UHV than delaminated regions exhibiting only smaller-scale crystallinity. This finding is particularly relevant as devices based on MoS\textsubscript{2}, and other TMDs, are scaled up to larger wafer-scale areas rather than the small exfoliated flakes that have been reported to exhibit self-cleaning. Further studies will be needed to determine whether the mechanism of contamination differs between areas with different levels of crystallinity.

Finally, temperature-programmed helium reflectivity measurements confirmed that adventitious carbon can be desorbed by mild annealing ($\sim\SIrange{80}{200}{\degreeCelsius}$) and allowed us to estimate the activation energy of desorption to be $\sim\SI{1}{eV}$.

Our findings confirm that UHV alone does not necessarily ensure clean TMD surfaces and highlight SHeM as a powerful tool for quantitative, non-invasive cleanness characterisation of such materials at the microscopic scale. Because helium scattering is chemically agnostic, our approach can easily be extended to other forms of lightly bound surface contamination. This enables quantitative monitoring of surface cleanness over extended areas with $\mu$m-scale spatial resolution — a capability critical for ensuring reproducibility in 2D material devices. Together, these features open a pathway for real-time monitoring of surface cleanness, supporting greater reliability for integration of 2D materials into future electronic and sensing applications.

\section{Methods}

\subsection{Sample preparation}
The bulk MoS\textsubscript{2} (natural crystal) used in B-SHeM (all data except figure \ref{fig:monolayer_mos2}) and XPS measurements was purchased from 2D Semiconductors Ltd. The sample was cleaved with scotch tape and stuck on the sample stub with silver DAG, before being installed in the B-SHeM sample chamber. This exfoliated bulk MoS\textsubscript{2} after the completion of all B-SHeM measurements was used in XPS measurements.

Monolayer MoS\textsubscript{2} for A-SHeM measurements (figure \ref{fig:monolayer_mos2}) was prepared by mechanically exfoliating a bulk MoS\textsubscript{2} crystal (2D Semiconductors Ltd.) and transferring it onto few-layer hBN supported on a SiO\textsubscript{2} substrate. Each monolayer had a lateral flake dimension of approximately $\SI{15}{\micro\metre}$.



\subsection{SHeM measurements}

\begin{figure}[t]
    \centering
    \includegraphics[width=0.78\linewidth]{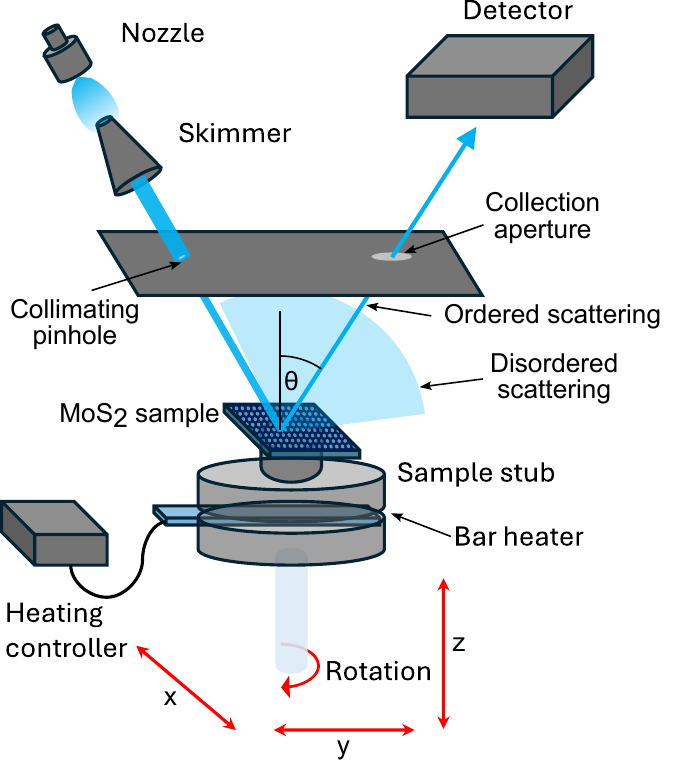}
    \caption{\textbf{A schematic depiction of the Cambridge B-SHeM as employed in the current work.} High pressure helium gas is expanded into a vacuum and repeatedly collimated to produce a microscopic probe. The helium atoms scatter from the sample either in ordered manner (specular/diffraction) or disordered manner (diffuse). The atoms scattered in a specific direction are collected and detected. The sample is manipulated in $xy$ to image and $yz$ to perform diffraction scans. A specialised, heater compatible, SEM sample stub is used to control the sample temperature.}
    \label{fig:schematic}
\end{figure}

A working schematic of the SHeM used for most of the current work (the Cambridge B-SHeM \cite{Zhao2025}) is shown in figure \ref{fig:schematic}(a). High pressure helium gas undergoes a supersonic expansion into vacuum \cite{Barr2012,Eder2018} through a $\SI{10}{\micro\metre}$ nozzle with the centreline selected by a $\SI{100}{\micro\metre}$ skimmer to produce a monochromatic beam. The beam is further collimated by a $\SI{30}{\micro\metre}$ pinhole mounted on a `pinhole plate' just before the sample. The sample is raster-scanned under the helium beam, and the scattered atoms collected via and aperture in the pinhole plate are counted by a mass-spec detector pixel-by-pixel to form an image. The temperature of the sample can be controlled using a specialised sample stub containing a bar heater, with a pair of thermocouple wires connected a proportional-integral-derivative controller. The sample can be moved in vacuum by piezoelectric motors, which provides three degrees of freedom in translational motions, and rotation about the $z$-axis. Measurements of diffraction, such as those presented in figures \ref{fig:clean_and_recontaminate}(e) and \ref{fig:monolayer_mos2}(c), are acquired by manipulating the position of the sample in $zy$ in order to change the detection angle $\theta$, while keeping the same spot on the sample under the beam.
Further details about the design of the instrument are given by Zhao et al. \cite{Zhao2025}, while details on the method to measure diffraction in SHeM can be found in von Jeinsen \textit{et al.} \cite{vonJeinsen2023}.

The exfoliated bulk MoS\textsubscript{2} was imaged at room temperature at the first instance, using the B-SHeM (roughly along the specular direction). The sample was then heated, and an image was taken at $\sim\SI{200}{\degreeCelsius}$ with the same instrument settings. The spot D on the flat MoS\textsubscript{2} surfaces was selected to measure diffraction (supplementary figure 3). The position and orientation of the sample to detect the specular peak along $\langle 110 \rangle$ were recognised. Subsequently, the sample was exposed to air for approximately half an hour and was placed in the vacuum chamber for about 24 hours, at room temperature. After this process, the sample surface at the specular peak position along $\langle 110 \rangle$ was imaged, and diffraction measurement along $\langle 110 \rangle$ was conducted at room temperature. The sample was heated again to take an image and measure the diffraction along $\langle 110 \rangle$ at $\sim\SI{200}{\degreeCelsius}$.

A flat region and a delaminated region of MoS\textsubscript{2} surfaces were selected to study microscopic contamination processes. After taking the images of these regions at $\sim\SI{200}{\degreeCelsius}$, the sample was naturally cooled down to room temperature and a series of images was then taken. A pair of images (a flat region and a delaminated region) took 1.35 h. Additionally, the diffraction from spot D along $\langle 110 \rangle$ was measured before taking every pair of images.

Starting with the MoS\textsubscript{2} surfaces at the contaminated state, the temperature programmed helium reflectivity (figure \ref{fig:desorption}(a)) was measured on spot D in figure \ref{fig:clean_and_recontaminate}, at the specular position along $\langle 110 \rangle$. During the experiment, the sample was heated from room temperature to $\sim\SI{300}{\degreeCelsius}$ with $\sim\SI{0.5}{\degreeCelsius/s}$. In addition, a flat region and a delaminated region were imaged at different temperatures from the contaminated state at room temperature, as shown in supplementary figure 8.

For the data in figure \ref{fig:monolayer_mos2} the Cambridge A-SHeM was used. Details of the original design can be found in ref. \cite{barr_design_2014}, and the modifications and experimental procedures for HAMD can be found in ref. \cite{vonJeinsen2023}. The same sample heater stub as used in the B-SHeM measurements, and shown diagrammatically in figure \ref{fig:schematic}, was used.

Monolayer MoS\textsubscript{2} was firstly imaged at room temperature using the A-SHeM. A selected spot on the MoS\textsubscript{2} surfaces was measured by varying $z$ corresponding to varying detection angles. Afterwards, the sample was heated to $\sim\SI{240}{\degreeCelsius}$. Another image of the monolayer MoS\textsubscript{2} surfaces and the in-plane diffraction along $\langle 110 \rangle$ from a microscopic spot were measured.


\section{Data availability}

All data will be made publicly available upon publication.
\FloatBarrier
\setlength{\bibsep}{0pt plus 0ex}
\input{main.bbl}


\section{Acknowledgements}

The work was supported by EPSRC grant EP/R008272/1. The work was performed in part at CORDE, the Collaborative R\&D Environment established to provide access to physics related facilities at the Cavendish Laboratory, University of Cambridge and EPSRC award EP/T00634X/1. S.M.L. acknowledges support from IAA award EP/X525686/1. The XPS data collection was supported by the Henry Royce Institute for advanced materials through the Equipment Access Scheme enabling access to the Royce XPS facility at Cambridge; Cambridge Royce Facilities grant EP/P024947/1 and Sir Henry Royce Institute – recurrent grant EP/R00661X/1. K.W. acknowledges financial support from the Cambridge Trust and CSC. We would like to thank Yiru Zhu at the Department of Material Science and Metallurgy, University of Cambridge, for preparation of the monolayer MoS\textsubscript{2} sample. We thank Paul Dastoor, Bill Allison, Yan Wang, and Han Yan for useful discussions. 

\section{Author contributions}
C.Z. and B.L. performed B-SHeM measurements; C.Z., B.L., and A.P.J. analysed the data. K.W. performed A-SHeM measurements. C.Z., B.L., and S.G. performed XPS measurements and data analysis. S.M.L. performed the ray tracing simulation. C.Z. and S.M.L. wrote the manuscript with contributions from all the authors. All authors contributed to discussion and data interpretation. B.L., A.P.J., and S.M.L. supervised the research.

\section{Competing interests}
The authors declare no competing interests.

\end{document}


\maketitle

\section{Identifying the contaminant as adventitious carbon}

\subsection{XPS measurements}

\begin{figure*}[t]
    \centering
    \includegraphics[width=0.9\linewidth]{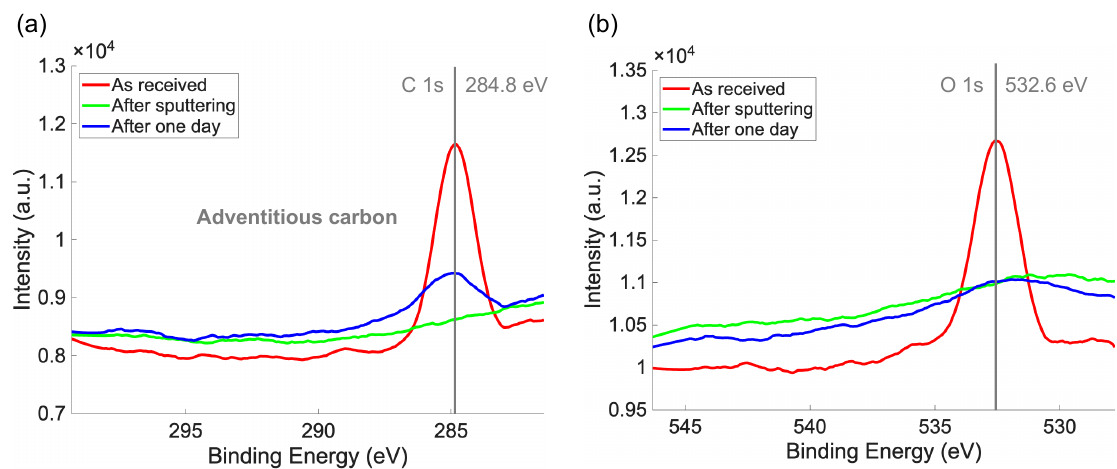}
    \refstepcounter{suppfigure}
    \caption*{\textbf{\thesuppfigure\ \textbar\ }XPS spectra of MoS\textsubscript{2} as received (red), after Ar ion sputtering (green) and after one day in the UHV sample chamber (blue). In (a), the peak of C 1s around 284.8 eV shows the existence of adventitious carbon on contaminated MoS\textsubscript{2} surfaces. In (b), the peak of O 1s around 532.6 eV illustrates the existence of oxygen.}
    \label{fig:XPS measurements}
\end{figure*}

To identify the chemical nature of the contamination, X-ray photoelectron spectroscopy (XPS) was conducted on the same bulk MoS\textsubscript{2} sample as that in the main text, using a Thermo Scientific Escalab 250Xi instrument fitted with a monochromated Al k$\alpha$ X-ray source (1486.7 eV) \cite{Greczynski2020}. The MoS\textsubscript{2} crystal was re-exposed to air prior to loading into the XPS UHV sample chamber (less than $10^{-8}$ mbar). Afterwards the sample was measured three times.
\begin{itemize}
    \item First, XPS measurement was performed straight after loading into the XPS UHV chamber, without heating or other treatments.
    \item Immediately after the previous set of XPS measurements, the MoS\textsubscript{2} crystal was cleaned by Ar\textsuperscript{+} sputtering. Then XPS data were taken.
    \item After the MoS\textsubscript{2} sample spent 24 hours in the UHV chamber at room temperature, XPS measurement was conducted again. There was no air exposure throughout the three sets of XPS experiments.
\end{itemize}
In the XPS measurements, the only noticeable peaks, apart from those belonging to Mo and S, are the C 1s and O 1s peaks. The C 1s spectra (\ref{fig:XPS measurements}(a)) demonstrate a distinct peak at $\sim$ 284.8 eV in the as-received MoS\textsubscript{2} sample (red), which disappears after Ar\textsuperscript{+} sputtering (green) and re-emerges with a lower intensity after one day in vacuum (blue). The C 1s signal at $\sim$ 284.8 eV suggests the presence of C–C/C–H bonds of adventitious carbon \cite{Grey2024,Greczynski2020,Greczynski2021,Greczynski2024}. In addition, the peak of O 1s  at $\sim$ 532.6 eV (\ref{fig:XPS measurements}(b)), corresponding to the binding energy of the C-OH/C-O-C bonds \cite{Giubileo2019}, is also expected for adventitious carbon since it contains approximately 25\% oxygen-bound carbon species \cite{Grey2024}. 
The O 1s peak is prominent in the as-received MoS\textsubscript{2}, and vanishes after sputtering and remains undetectable after one day in UHV. The XPS measurements confirm that the adsorbates on MoS\textsubscript{2} surfaces are adventitious carbon, which can be removed by sputtering but gradually reappears even under UHV. Similar findings were reported by Addou \textit{et al.}, where XPS measurements on exfoliated MoS\textsubscript{2} revealed prominent C 1s and O 1s peaks, and contaminants desorbed after annealing in UHV at $\SI{400}{\degreeCelsius}$ for 15 minutes \cite{Addou2015}.

All the XPS spectra were measured at room temperature with an X-ray beam size of $\SI{650}{\micro\metre}$ and a pass energy of 20 eV at a step size of 0.1 eV. Electronic charge neutralisation was achieved using an ion source (ion gun current = $\SI{100}{\micro\ampere}$, ion gun voltage = 40 V). In all the XPS spectrum measurements, the same sample region was measured with the same instrument settings. The Ar\textsuperscript{+} sputtering was performed with a 2 keV monatomic source for 200 seconds at the rate of 0.49 nm/s (ref. Ta\textsubscript{2}O\textsubscript{5}). The beam size of sputtering was 1 mm.
 
\subsection{Desorption and adsorption measurements}

To further verify that the contamination observed by SHeM is adventitious carbon but not other common adsorbates on crystal surfaces, temperature-programmed desorption (TPD) experiments and adsorption experiments were conducted on the MoS\textsubscript{2} sample in the Cambridge B-SHeM vacuum chamber \cite{Zhao2025}.

In a TPD experiment, the MoS\textsubscript{2} sample is heated up from room temperature to over $\SI{300}{\degreeCelsius}$ in UHV, while a Hiden HAL/3F RC301 PIC300 mass spectrometer is used to monitor the amount of desorbed atoms and molecules. Several masses (for example, 1, 2, 16, 18, 28, 30, 32, 44, 46, and 58 AMU) were measured, which correspond to common surface adsorbates in UHV, such as hydrogen, oxygen, water, carbon monoxide, nitrogen, carbon dioxide, etc. If any of the species monitored is adsorbed on the surface, as temperature rises, the mass spec signal will go up rapidly (corresponding to the start of desorption) and then down (corresponding to all adsorbates desorbing and pumped away). However, in all the TPD experiments, as the temperature rose, the mass spectrometer signal did not show a peak as a function of temperature or time. Instead, the signal went up very slightly and stayed constant. Afterwards, the MoS\textsubscript{2} was removed from the vacuum chamber, and the same TPD experiments were performed; however there is no evident difference from the experiments with MoS\textsubscript{2} in the chamber. This indicates that there was only a small amount of desorption from the sample stage, substrate, etc., but there was no desorption of the monitored species from MoS\textsubscript{2}. Therefore, the increase in helium reflectivity as MoS\textsubscript{2} was heated (see figures 2, 3, and 5 in the main paper) is not due to desorption of common simple molecule contaminants that leak in from the atmosphere.

In the adsorption experiments, the helium reflectivity of MoS\textsubscript{2} at specular scattering condition was monitored as a gas was introduced into the vacuum chamber through a leak valve. The experiment was repeated with the a different gas each time, including hydrogen, carbon dioxide, oxygen, water vapour, and air. The pressure of each gas was over $10^{-6}$ mbar and the exposure time was more than one hour. However, in all the experiments, there was no noticeable change in helium reflectivity. Therefore, these molecules were excluded as contaminants on the sample surface when the sample gets contaminated in vacuum, as shown in figure 4 in the main paper.

Both TPD experiments and adsorption experiments further confirm that the observed change in helium signal in the main paper was not induced by adsorption/desorption of simple molecules, but by adventitious carbon.

\section{Supplementary SHeM measurements and analysis}

\subsection{Diffraction scans at different levels of adventitious carbon contamination}

\begin{figure}[t]
    \centering
    \includegraphics[width=0.4\linewidth]{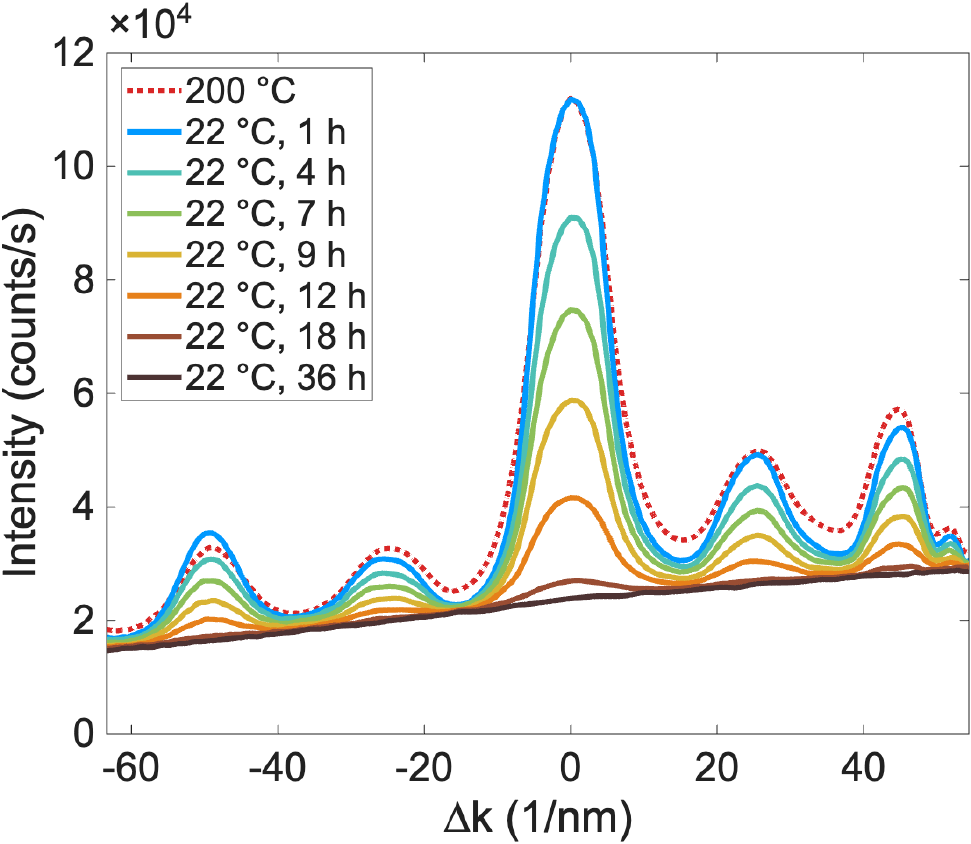}
    \refstepcounter{suppfigure}
    \caption*{\textbf{\thesuppfigure\ \textbar\ }Helium atom micro diffraction scans taken from the spot D (see figure 2 of the main text) of MoS\textsubscript{2} surfaces, first at $\SI{200}{\degreeCelsius}$ then at various intervals after cooling to $\SI{22}{\degreeCelsius}$.}
    \label{Sfig:Diffraction during contamination}
\end{figure}

Helium atom micro diffraction scans \cite{vonJeinsen2023} were taken from the sample at different times during the recontamination process presented in figure 4 of the main text. The point measured was spot D, and the full results are presented in \ref{Sfig:Diffraction during contamination}. An initial diffraction scan (red dotted line) was acquired at $\SI{200}{\degreeCelsius}$. Subsequently, the sample was cooled to room temperature in the UHV chamber. A series of diffraction scans were taken after various time intervals following the cooling. Gradual suppression of diffraction over time at room temperature is evident, following the same trends observed in figure 2 of the main text and confirms our interpretation of the reflectivity measurements and image contrast: we observe a progressive loss of atomic-scale surface order resulting from increased adsorption of the adventitious carbon.

\subsection{2D diffraction measurement}

\begin{figure}[t]
    \centering
    \includegraphics[width=0.64\linewidth]{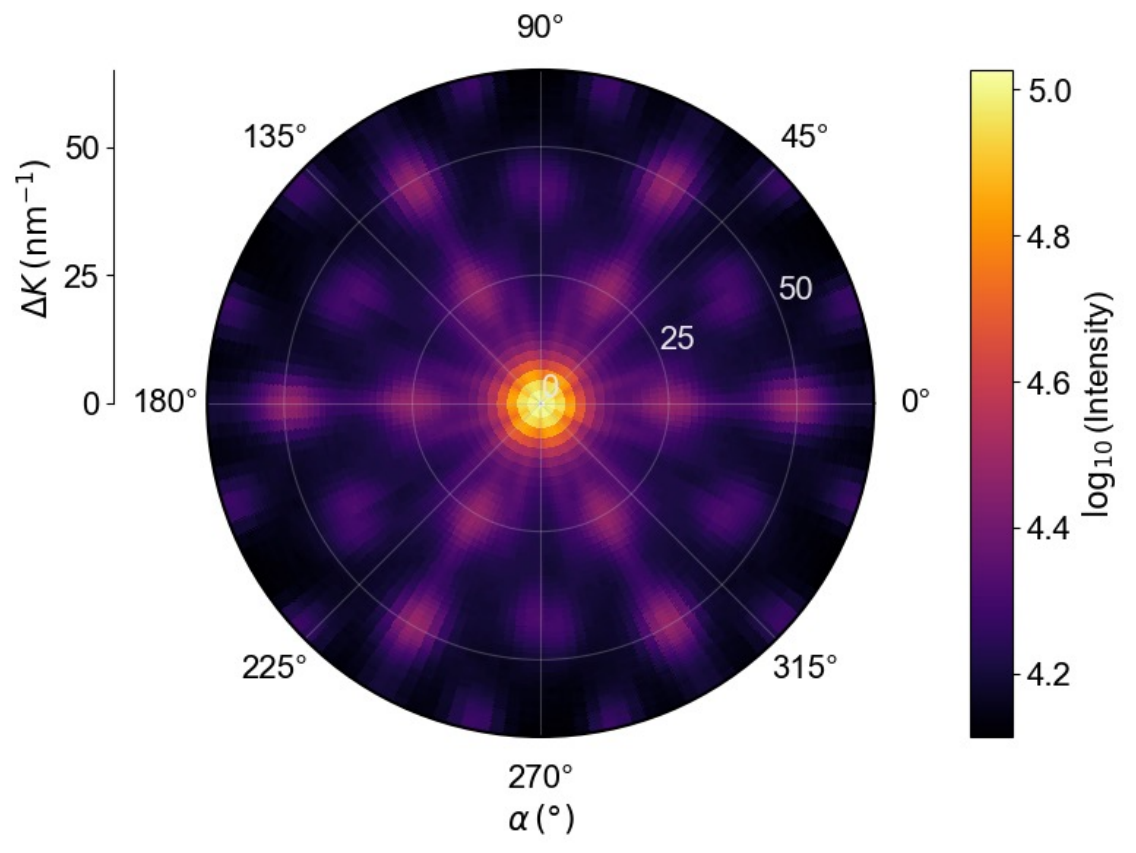}
    \refstepcounter{suppfigure}
    \caption*{\textbf{\thesuppfigure\ \textbar\ }2D diffraction measurements from the spot D (see figure 2 in the main text) of MoS\textsubscript{2} surfaces. The hexagonal lattice can be observed clearly. The specular spot in the middle has the highest intensity, and the diffraction spots along $\langle 110 \rangle$ display a higher intensity than those along $\langle 100 \rangle$.}
    \label{Sfig:2D diffraction}
\end{figure}

\ref{Sfig:2D diffraction} shows 2D diffraction measurements \cite{vonJeinsen2023} from MoS\textsubscript{2} crystal surfaces using the B-SHeM. In the measurement, the sample was moved in such a way that the same spot on the sample (spot D in figure 2) was illuminated by the incoming He beam. The translational motions in $x$, $y$, and $z$ directions change $\theta$ (figure 6 in main text), and thus control the magnitude of parallel momentum transfer, $\Delta K$; while the rotation along $z$ controls the direction of the parallel momentum transfer. In experiments, $\theta$ was only rotated by $60^\circ$ and the rest of the figure is created using symmetry. The data was taken with a sample temperature of $\SI{200}{\degreeCelsius}$.

\subsection{Diffraction from delaminated regions at different levels of contamination}

\begin{figure*}[htbp]
    \centering
    \includegraphics[width=0.78\linewidth]{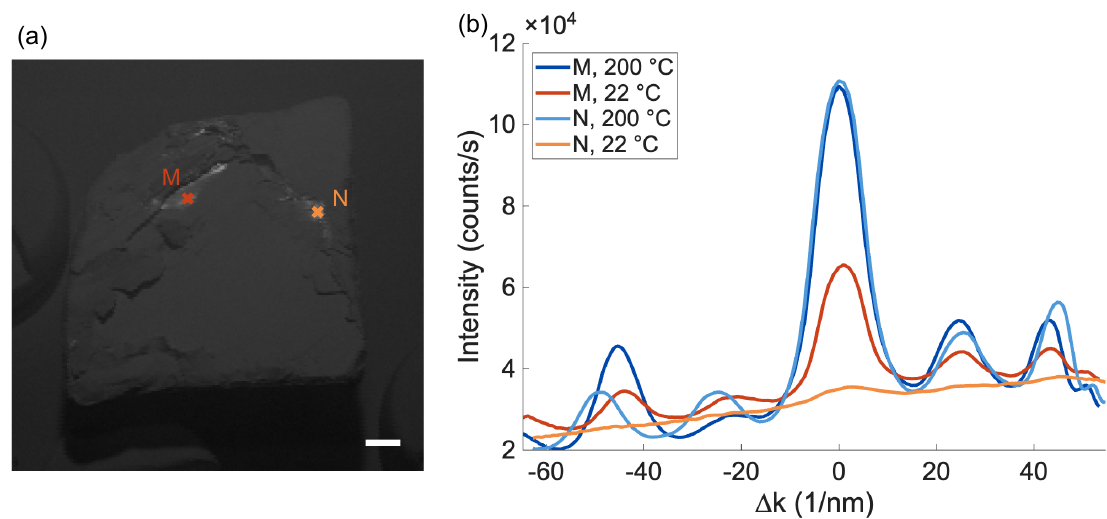}
    \refstepcounter{suppfigure}
    \caption*{\textbf{\thesuppfigure\ \textbar\ }(a) Recontaminated MoS\textsubscript{2} surfaces with small relative clean region. Scale bar, $\SI{500}{\micro\metre}$. (b) Diffraction measured from the spots M and N at room temperature and $\SI{200}{\degreeCelsius}$. The bright region around delaminated surfaces is clean after venting the chamber for 30 minutes and standing in UHV for 24 hours at room temperature, showing diffraction from MoS\textsubscript{2} surfaces.}
    \label{Sfig:Diffraction from bright spot}
\end{figure*}

In figure 2(c) in the main text, the MoS\textsubscript{2} was mostly contaminated apart from a few regions which appeared bright. To confirm that the small bright regions was due to MoS\textsubscript{2} being partially contaminated (figure 1(b) in the main text), diffraction scans were taken on two spots, as shown in \ref{Sfig:Diffraction from bright spot}. The results show that on the two spots, the contaminated sample at room temperature showed weakened diffraction compared to clean samples at $\SI{200}{\degreeCelsius}$. The discrepancy in the positions of the diffractions peaks is caused by the slight tilt in the measured region of the sample.

\subsection{Scattering distribution from contaminated MoS\textsubscript{2}}

\begin{figure}[]
    \centering
    \includegraphics[width=0.48\linewidth]{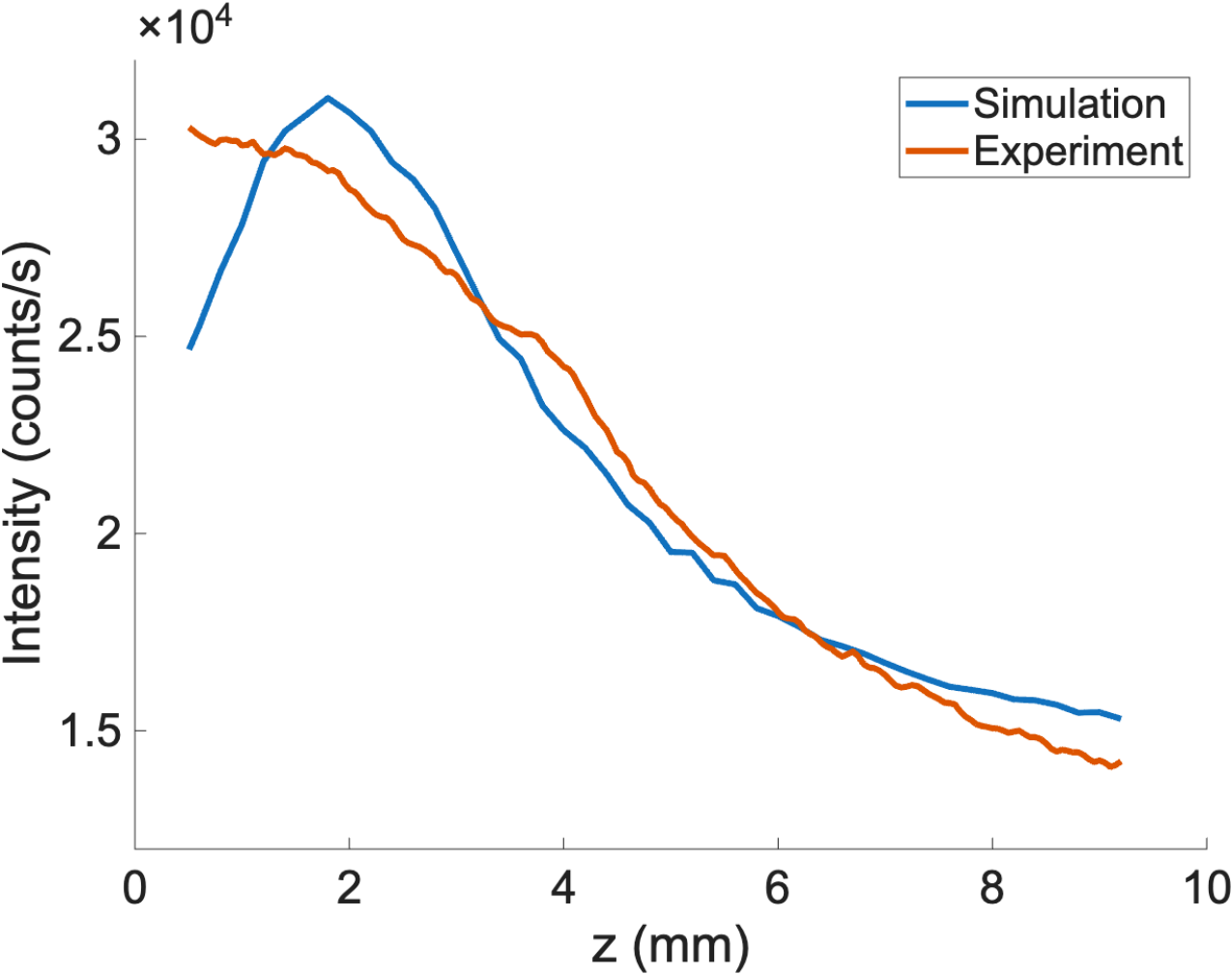}
    \refstepcounter{suppfigure}
    \caption*{\textbf{\thesuppfigure\ \textbar\ }The experimental and simulated intensity from the MoS\textsubscript{2} surfaces covered with the adventitious carbon as a function of $z$. The simulation is performed with a cosine scattering distribution by ray tracing. The simulated scattering intensities consist of the direct scattering \& the multiple scattering of the direct beam and the diffused beam. The experimental data fit with the simulation at $z > 2$ mm, which confirms that the scattered helium atoms from the contaminated crystalline surfaces follow a cosine-like distribution. The experimental data at $z < 2$ mm does not match the simulation because of the unexpected high background gas at a short pinhole to sample distance.}
    \label{Sfig:Scattering distribution}
\end{figure}

It has been previously demonstrated that diffuse scattering usually can be quantitatively model by a cosine scattering distribution \cite{Lambrick2022}. To investigate whether the scattering from MoS\textsubscript{2} follow cosine scattering, the helium signal was measured as a function of $z$ on a contaminated surface. Afterwards, a ray tracing simulation was carried out that compares with experiments. The results of both the experiment and simulation are plotted in \ref{Sfig:Scattering distribution}. The overall agreement between the experiment and the simulation shows that the scattering distribution from contaminated surface is cosine.

\subsection{Complete SHeM image dataset of adventitious carbon contamination and removal in UHV}

\begin{figure*}[t]
    \centering
    \includegraphics[width=1.0\linewidth]{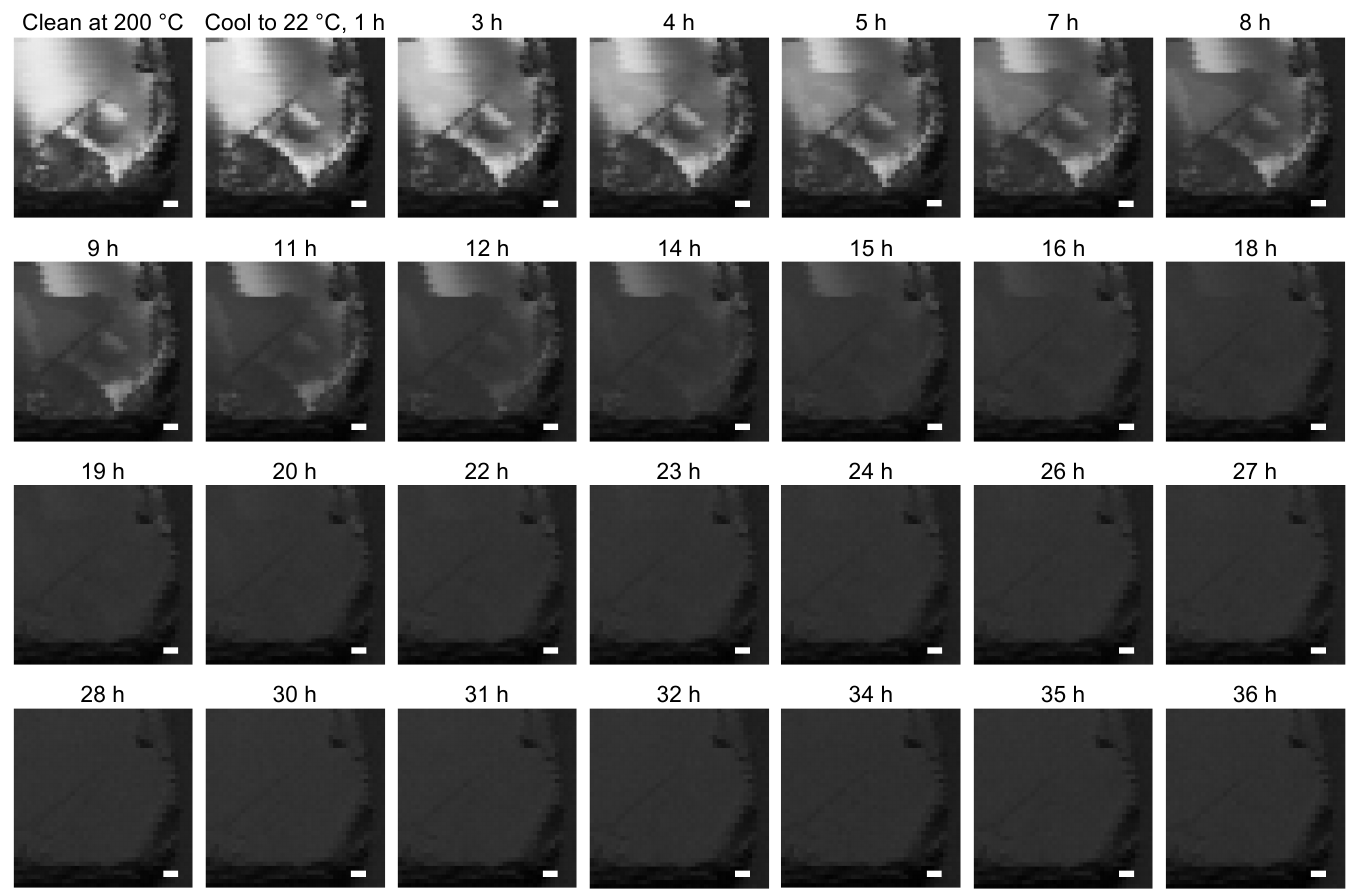}
    \refstepcounter{suppfigure}
    \caption*{\textbf{\thesuppfigure\ \textbar\ }Microscopic contamination process of the flat MoS\textsubscript{2} surfaces. Scale bar, $\SI{100}{\micro\metre}$.}
    \label{Sfig:Adsorption_flat_all}
\end{figure*}

\begin{figure*}[]
    \centering
    \includegraphics[width=1.0\linewidth]{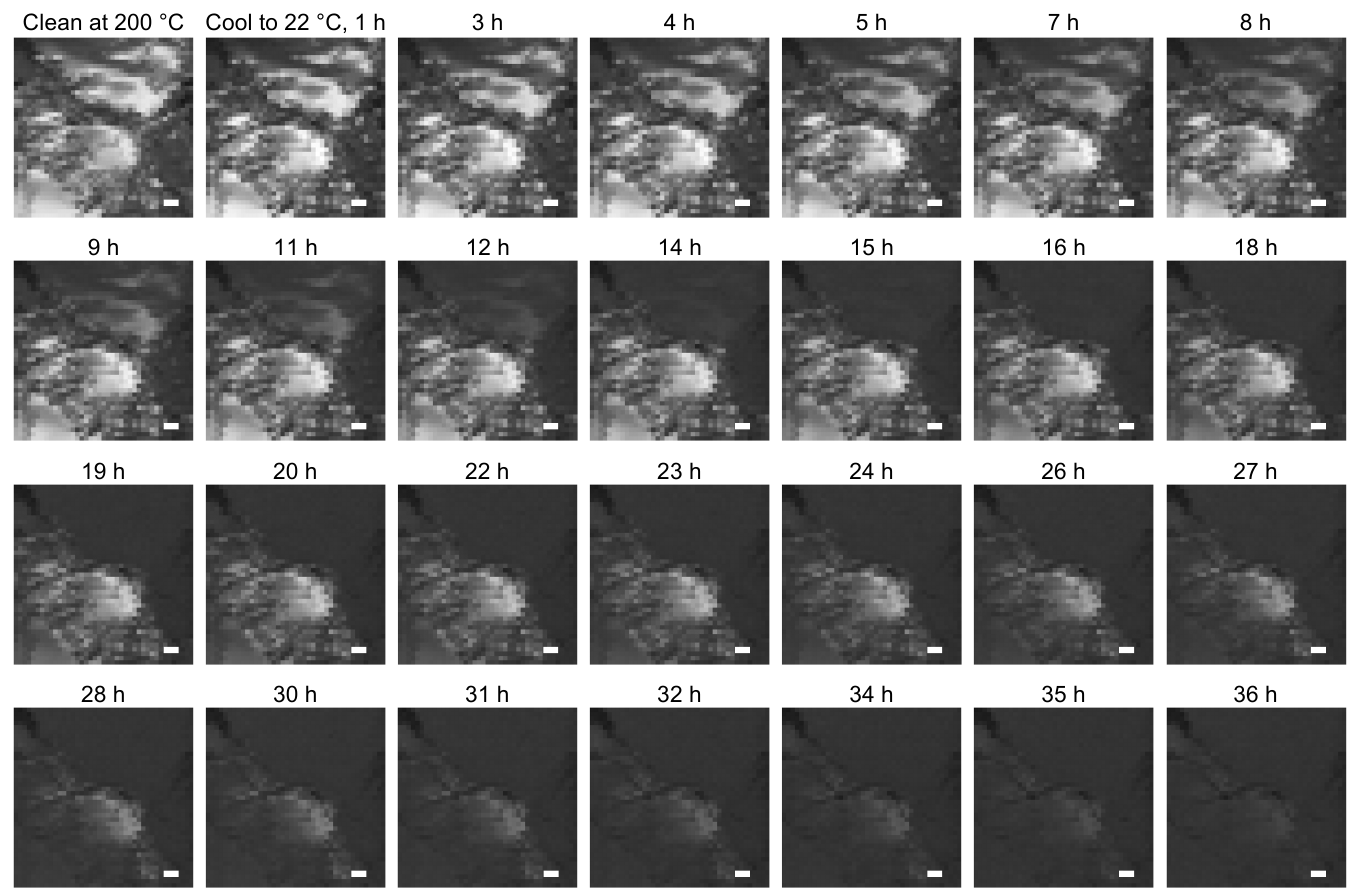}
    \refstepcounter{suppfigure}
    \caption*{\textbf{\thesuppfigure\ \textbar\ }Microscopic contamination process of the delaminated MoS\textsubscript{2} surfaces. Scale bar, $\SI{100}{\micro\metre}$.}
    \label{Sfig:Adsorption_layered_all}
\end{figure*}

When measuring the contamination process of the MoS\textsubscript{2} in real space (figure 4 in main text), over 20 SHeM micrographs were taken, with about 1.35 hours between them. \ref{Sfig:Adsorption_flat_all} shows the all the SHeM micrographs of the flat region. Five of them are displayed in figure 4(a)-(e). The subfigures in \ref{Sfig:Adsorption_layered_all} are those corresponding to figure 4(g)-(k). Two videos corresponding to the contamination process are attached.

\begin{figure*}[]
    \centering
    \includegraphics[width=1.0\linewidth]{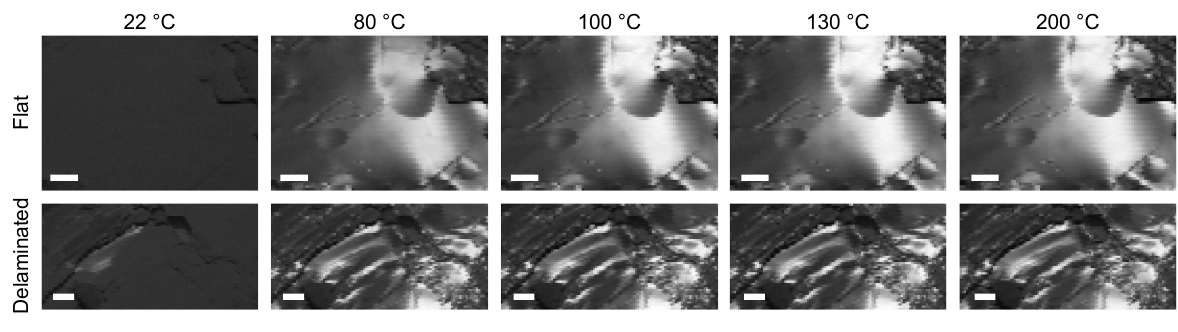}
    \refstepcounter{suppfigure}
    \caption*{\textbf{\thesuppfigure\ \textbar\ }SHeM micrographs of the flat and delaminated MoS\textsubscript{2} surfaces at different temperature. Scale bar, $\SI{300}{\micro\metre}$.}
    \label{Sfig:Desorption_all}
\end{figure*}

In figure 5 of the main text, SHeM micrographs were shown at three different temperatures, which are $\SI{22}{\degreeCelsius}$, $\SI{80}{\degreeCelsius}$, and $\SI{200}{\degreeCelsius}$. \ref{Sfig:Desorption_all} shows two more micrographs at $\SI{100}{\degreeCelsius}$ and $\SI{130}{\degreeCelsius}$, of both delaminated and flat regions.

\subsection{Redhead model}

\begin{figure*}[t]
    \centering
    \includegraphics[width=0.5\linewidth]{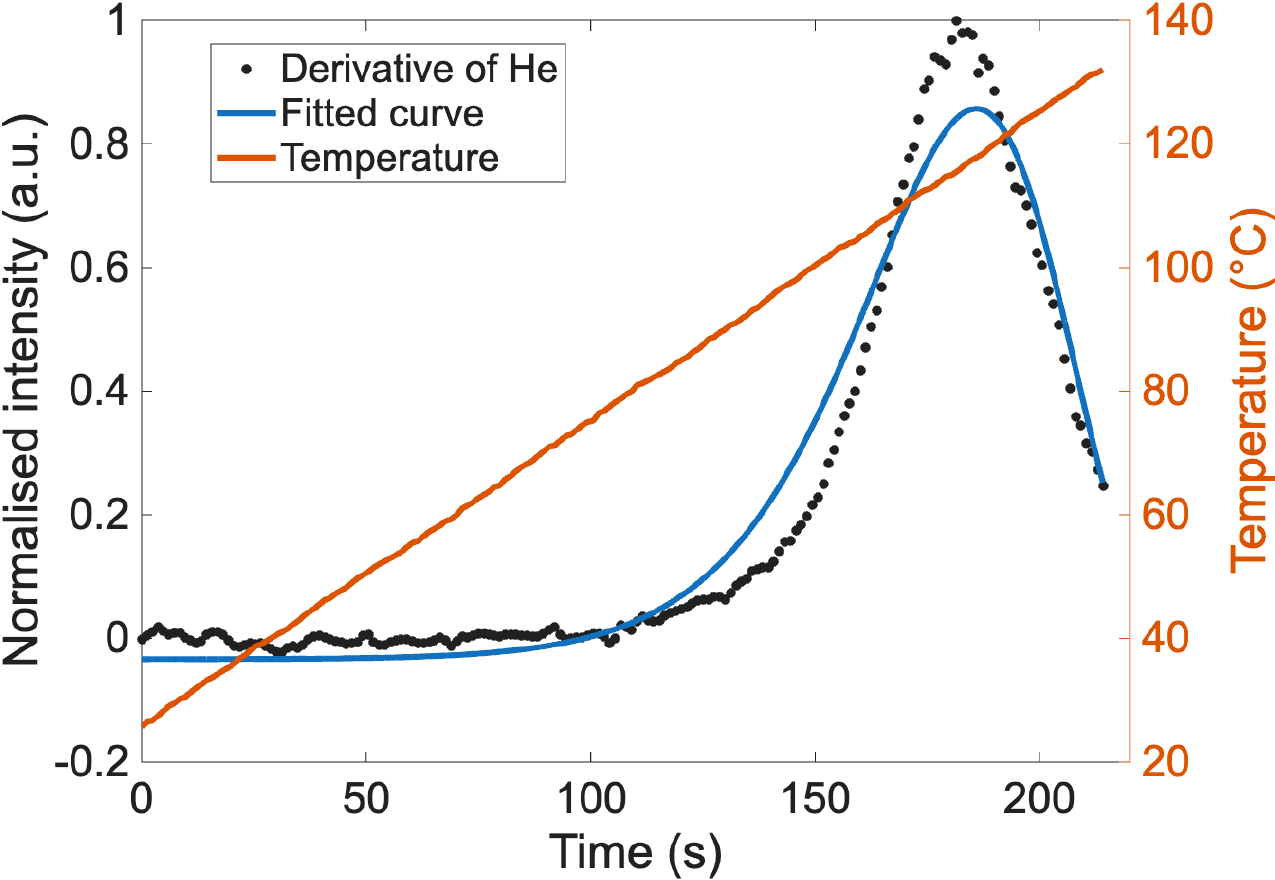}
    \refstepcounter{suppfigure}
    \caption*{\textbf{\thesuppfigure\ \textbar\ }The derivative of helium reflectivity fitted with the first order Redhead model. This model roughly estimates the activation energy of the adventitious carbon since the derivative of helium is similar to TPD measurements. Traditional TPD measurements failed to resolve the peaks for a tiny and complex mixture such as adventitious carbon. This temperature-programmed helium reflectivity method is a supplement to traditional TPD measurements. This method can also provide spatial resolution to characterise different surface regions of the sample.}
    \label{Sfig:TPD fitting}
\end{figure*}

\ref{Sfig:TPD fitting} shows the same data in figure 5(a) of the main text, but in the time domain. The sample temperature increases linearly with time. So the Redhead model \cite{Redhead1962} can be used, which is written as
\begin{equation}
r(t) = -\frac{d\theta}{dt} = \nu_0 \theta^n \exp\left(-\frac{E_{\text{a}}}{R(T_0+Bt)}\right),
\end{equation}
where $r$ is the desorption rate, $\theta$ is the coverage of the adsorbates, $t$ is time, $\nu_0$ is the frequency factor, $n$ is the order of desorption, $E_\text{a}$ is the activation energy, $R$ is the gas constant, $T_0$ is the initial temperature and $B$ is the heating rate. This equation with the first order ($n$=1) can be calculated, and the measured desorption intensity is expressed as,
\begin{equation}
I(t) = I_\text{bg} + C \nu_0 
\exp\!\left[-\frac{E_\text{a}}{R(T_0 + Bt)}\right]
\exp\!\left[
-\frac{\nu_0}{RB}
\left(
\exp\!\left[-\frac{E_\text{a}}{R(T_0 + Bt)}\right]
R(Bt + T_0)
+ E_\text{a}\mathrm{Ei}\!\left(-\frac{E_\text{a}}{R(T_0 + Bt)}\right)
\right)
\right].
\end{equation}
$I_\text{bg}$ is the measured background intensity; $C$ is a constant and $\mathrm{Ei}$ is the exponential integral function. The fitted results are $C = 13.4 \pm 0.4$, $\nu_0 = (6.3\pm 1.5)\times10^{13}\ \SI{}{s^{-1}}$, $E_\text{a} = 1.13 \pm 0.01\  \SI{}{eV}$, and $I_\text{bg} = -0.03 \pm 0.01$.

\setlength{\bibsep}{0pt plus 0ex}
\input{Supplementary_information.bbl}

%% file: main.bbl
\providecommand{\latin}[1]{#1}
\makeatletter
\providecommand{\doi}
  {\begingroup\let\do\@makeother\dospecials
  \catcode`\{=1 \catcode`\}=2 \doi@aux}
\providecommand{\doi@aux}[1]{\endgroup\texttt{#1}}
\makeatother
\providecommand*\mcitethebibliography{\thebibliography}
\csname @ifundefined\endcsname{endmcitethebibliography}  {\let\endmcitethebibliography\endthebibliography}{}

%% file: Supplementary_information.bbl
\providecommand{\latin}[1]{#1}
\makeatletter
\providecommand{\doi}
  {\begingroup\let\do\@makeother\dospecials
  \catcode`\{=1 \catcode`\}=2 \doi@aux}
\providecommand{\doi@aux}[1]{\endgroup\texttt{#1}}
\makeatother
\providecommand*\mcitethebibliography{\thebibliography}
\csname @ifundefined\endcsname{endmcitethebibliography}  {\let\endmcitethebibliography\endthebibliography}{}